# Femtosecond Photocurrents at the Pt/FeRh Interface

R. Medapalli[1,2], G. Li[3], Sheena K. K. Patel[1], R. V. Mikhaylovskiy[3,4], Th. Rasing[3], A. V. Kimel[3], and E. E. Fullerton[1]

[1]Center for Memory and Recording Research, University of California, San Diego, La Jolla, California 92093-0401, USA

[2]Department of Physics, School of Sciences, National Institute of Technology, Andhra Pradesh-534102, India

[3]Radboud University, Institute for Molecules and Materials, Heyendaalseweg 135, Nijmegen, The Netherlands

[4]Department of Physics, Lancaster University, Bailrigg, Lancaster LA1 4YW, UK

**Femtosecond laser excitation of FeRh/Pt bilayers launches an ultrafast pulse of electric photocurrent in the Pt-layer and thus results in emission of electromagnetic radiation in the THz spectral range. Analysis of the THz emission as a function of polarization of the femtosecond laser pulse, external magnetic field, sample temperature and sample orientation shows that photocurrent can emerge due to vertical spin pumping and photo-induced inverse spin-orbit torque at the FeRh/Pt interface. The vertical spin pumping from FeRh to Pt does not depend on the polarization of light and originates from ultrafast laser-induced demagnetization of the ferromagnetic phase of FeRh. The photo-induced inverse spin-orbit torque at the FeRh/Pt interface can be described in terms of a helicity-dependent effect of circularly polarized light on the magnetization of the ferromagnetic FeRh and subsequent generation of a photocurrent.**

In recent years, research indicates antiferromagnetic (AF) materials can dramatically improve spintronics technology in terms of density and speed [1]. Unlike ferromagnets, antiferromagnetic materials produce either no or very small stray fields that would minimize cross-talking between neighboring devices, and have up to $10^3$ higher frequency spin resonances [2], which promises to push the operation frequency of spintronics devices to the THz domain [3]. This is why understanding spin transport in antiferromagnets is emerging as one of the hottest topics in magnetism research.

FeRh is a rather unique AF material which undergoes a first order phase transition to ferromagnetic (FM) state upon temperature increase. Due to this property, FeRh offers a unique playground to investigate of spintronics phenomena across the transition between AF and FM phases in the same material [4,5]. For instance, vertical and lateral spin pumping during the phase transition have been recently reported for Pt/FeRh and Py/FeRh bilayers [6]. Evidence of THz vertical spin currents in Pt/FeRh were obtained with the help of THz emission spectroscopy in Ref. [7]. Although, several hypotheses inspired by Refs [8,9] have been suggested, the exact origin of the spincurrents was not identified. One of the reasons for the difficulties hampering interpretation was the observation of rather different temperature dependencies of the net magnetization, measured with the help of a vibrating sample magnetometer, and the amplitude of the laser-induced THz emission. While the magnetization measurements revealed a temperature hysteresis typical for first-order phase transitions, such a hysteresis was not seen in the THz measurements.

Here, we reveal spin-dependent sources of laser-induced THz emission in a FeRh/Pt bilayer. The most intense source does not depend on the polarization of light (helicity-independent), this is evidenced by the THz vertical spin currents and it scales with the amount of the ferromagnetic (FM) phase and originates from ultrafast laser-induced demagnetization of ferromagnetic FeRh. The second, weaker source of the THz emission depends on the polarization of light due to photo-induced tilt of the magnetization and inverse spin-orbit torque effect at Pt/FeRh interface.

A 40 nm thick FeRh thin film was deposited via sputter deposition onto a MgO(001) single crystal and a capping layer of 5 nm thick Pt was added. We chose the thickness of Pt such that the inverse spin-Hall effect (ISHE) based spin-to-charge conversion is more efficient [10,11]. The FeRh was deposited at 450 °C and then post annealed at 800 °C for 45 min to improve the crystalline and chemical order. The Pt layer was deposited at room temperature. The layer thicknesses are found from X-ray reflectometry (see Fig. 1a). The observed small period oscillations originate from the 40 nm FeRh layer, and the large period oscillations from the 5 nm Pt layer. The high-quality reflectometry (Fig.1a) and the x-ray diffraction (see Fig. 1b) indicate a smooth film with epitaxial growth of FeRh(001). The presence of FeRh (001) and (003) peaks indicate the chemical ordering in the AF-phase. At 300 K, the lattice along the normal to the film is relaxed leading to compressive strain determined by the MgO substrate that further lead to difference of expansion between the in-plane and out-of-plane directions at the AF-FM phase transition [12].

The AF-FM magnetic phase transition (Fig.1c) was probed by vibrating sample magnetometry (Fig.1d) showing a clear latency in the phase-transition with temperature, reflecting the first order magnetic phase transition. The transition temperature is 370 K under modest external magnetic field with a 20 K latency heat in hysteresis (Fig.1d). Strain and structurally inhomogeneous defects can shift and broaden the temperature of the phase transition, leading to co-existence of AF and FM phases in a broad range around the transition temperature. Such a co-existence of phases is typical for all first-order phase transitions. In addition, several groups reported observation of FM-domains at the MgO/FeRh interface at T=300 K, well below the temperature of the phase transition [12,13,14].

Our experimental approach to phot-excite the FeRh/Pt sample is similar to the one employed in Ref.[7], but included the possibility to change the external magnetic field and apply larger laser fluences (up to 9.5 mJ/cm$^2$). The schematic of the THz electric field emission spectroscopy combined with the ultrashort optical excitation is shown in Suppl. Fig.1. The 50-fs optical pulses with a central wavelength of 800 nm, were generated by a pulsed Ti:Sapphire laser with a regenerative amplifier operating at a repetition rate of 1 kHz. A pump pulse was focused onto the sample at normal incidence (**z**-axis in Fig.1) with a 2-mm-diameter spot size. The polarization of the laser pulse was varied between linear, right- and left-circular polarization. The magnetization is parallel to the **x**-axis, and we detect the $\hat{x}$, and $\hat{y}$, components of the electric field of the emitted THz radiation propagating along the **z**-axis. An external magnetic field was applied along the **x**-axis in the plane of the sample to control the magnetization direction. Two opposite directions of the applied external magnetic field along and anti-parallel to **x**-axis are denoted as $M^\uparrow$ and $M^\downarrow$, respectively. The THz emission generated from the sample was focused onto a ZnTe crystal by two gold coated parabolic mirrors. The THz electric field was measured by detecting the ellipticity changes of the gating pulse induced by the Pockel's effect in ZnTe using a balance detector in combination with a lock-in amplifier.

Figure 1e (left panel) illustrates vertical laser-induced spincurrents and their conversion into charge currents via the inverse spin Hall effect, which produces THz radiation with the electric field transients as depicted in Fig. 1e. This mechanism arises from ultrafast demagnetization of the sample and is independent of the helicity of the light. The right image in Fig. 1e is THz emission that depends on the circular polarization of the light. The THz emission arises form helicity-dependent femtosecond laser-induced rotation of the magnetization in the plane of the sample which, due to the inverse spin-orbit torque, generates a femtosecond pulse of electric current at the interface. We made use of wire-grid polarizers to separately measure the terahertz emission polarized along the **x** and **y** axes. This gives the freedom of differentiating the electric field components of helicity-independent ($E_y$) and –dependent ($E_x$) mechanisms. A thermocouple heater was used to vary temperature of the sample between 300-450K (heating process) or 450-300 K (cooling process).

The time traces of the helicity-independent and helicity-dependent THz electric field components measured at various temperatures are shown in Fig.2a and 2b, respectively. The waveforms corresponding to the dynamics of **y**-component (helicity-independent) exhibit identical shape with a monotonous change in the amplitude with heating. This suggests that spectrum of the THz emission and sub-picosecond dynamics of the helicity-independent spincurrent does not depend on the relative AF and FM phase contribution. The helicity-dependent electric field component is determined from four combinations of (σ, M),

$E_{X,odd} = (E_x^{(M\uparrow, \sigma+)} - E_x^{(M\uparrow, \sigma-)} - E_x^{(M\downarrow, \sigma+)} + E_x^{(M\downarrow, \sigma-)})/4$, and its corresponding dynamics measured at various temperatures are shown in Fig.2b. The time-traces of y-component of the pump induced THz electric field emission for two opposite directions of the applied external magnetic field (B = ±150 mT) measured at 300 K (AF-phase) are shown in Supplementary Fig.2a. The THz frequency ranges up to 3 THz and peaks at 0.75 THz (not shown). There is a change of sign in the detected THz waveforms upon reversal of the external magnetic field direction. We also pumped the sample from both the substrate-side as well as the Pt-capping side. The corresponding time traces of the THz emission signal are shown in Supplementary Fig.2b. For a given magnetic field direction, a change in sign of the laser-induced THz-signals after rotating the sample by 180 deg. Indicates that the emitter of the THz $E_Y$-field is of electric dipole origin [11,25]. The symmetries of the experiments are in full agreement with the mechanism discovered for Pt/Co bilayers [15], where a sub-picosecond demagnetization of Co launched a vertical spin current across Pt/Co interface and the Pt-layer converted the spin current into a charge current via the ISHE [16]. In our case, the spincurrent can originate either from magnetization induced in the AF-phase due to an ultrafast phase transition to the FM state [17,18,19], or due to an ultrafast demagnetization of the FM-domains [20] present in FeRh at higher temperatures or even below the temperature of the phase transition.

The peak value of the detected THz $E_y$-field components denoted as Max.($|E_Y|$) is plotted as a function of temperature for the heating and cooling processes (see Fig.3). For comparison, the static net magnetic moment curves (for both heating and cooling) are shown as dashed lines in Fig.3. Unlike the behavior of the net magnetization, similarly to Ref.[7], the electric field components do not show any temperature hysteresis. The temperature dependency of the peak value of THz $E_{x,odd}$-component, Max.($|E_{x,odd}|$) is also shown in Fig.3, exhibiting a trend similar to Max.($|E_y|$). Figure 4a shows the Max.($|E_y|$) as a function of the applied magnetic field strength at four different initial temperatures (300, 360, 370, 380, and 450 K). It is seen that the hysteresis becomes broader upon temperature increase. The Fourier spectrum of the emitted THz $E_Y$-radiation (see Fig. 4b) does not depend on the applied magnetic field, which shows that the field does not affect the sub-picosecond dynamics of the vertical spin currents across the interface.

It is known that a moderate magnetic field and a relatively narrow range of temperature change should not affect the timescale of ultrafast laser-induced demagnetization of metallic ferromagnets [22]. At the same time, launching a phase transition from an AF to a FM phase will generate ferromagnetic nuclei with random orientations [21] of their magnetizations and the speed with which the magnetizations will be aligned along the applied magnetic field should be a function of the magnetic field and temperature [21,23,24]. Therefore, the fact that the THz spectra are neither influenced by temperature nor magnetic field suggests that the origin of the THz emission is the vertical spincurrent across the FeRh/Pt interface, due to an ultrafast demagnetization of FeRh.

The lack of temperature hysteresis in the THz measurements (see Fig.3) is a result of the stroboscopic nature of the experiment, which implies measurement of the THz electric field averaging over many heating-cooling cycles. From the very beginning of ultrafast magnetism, it is well established that stroboscopic measurements of spin dynamics cannot adequately reveal a hysteresis [26]. More particularly, a temperature hysteresis implies that in a certain temperature range the medium has multiple stable states. The hysteresis can disappear in stroboscopic measurements, because after the very first pump-induced heating and cooling down every next pump will steer the medium along the cooling branch of the hysteresis. In this case, the dynamics will not depend on how the medium was brought in the very first initial state. Moreover, there is a finite probability that after each pump-induced heating the medium relaxes to one or another metastable state. The temperature dependence of the THz emission shown in Fig. 3 would mean that the probability of the relaxation to each of the two stable states is 50%.

Changing the polarization of the detected THz electric field, our setup becomes sensitive to a different source of the radiation. Supplementary Figure 3a shows the time traces of the THz electric field along the **x**-axis ($E_x$) obtained at $T =$ 450 K for the cases corresponding to pumping the sample from the side of the Pt-layer or the MgO-substrate, respectively. The sign change shows that the source of the radiation is of electric dipole origin. Changing the helicity of the pump or changing the polarity of the applied magnetic field also results in helicity dependence of the $E_x$-field (see Suppl. Fig.3b,c). The symmetry of the experiments is in agreement with the mechanism discovered for Pt/Co bilayers [9], where circularly polarized light exerted a torque on the magnetization of Co due to the inverse Faraday effect [27] thereby tilting it in the sample plane. This tilt resulted in an electric current at the Pt/Co interface due to the inverse spin-orbit torque effect [28,29]. The peak values of the THz $E_x$–field extracted from the curves shown in Fig.2b, are also plotted in Fig.3. It is seen that the $E_x$-field follows the same dependence as the $E_Y$-field showing the lack of hysteresis and revealing the fact that photo-induced spincurrents at the Pt/FeRh interface increases with the growth of the FM phase in FeRh. Finally, the Fourier spectra of the $E_x$ and $E_y$ electric fields (for T= 450 K, and B = 150 mT) are presented in Fig. 4c. The shift in the peak THz frequencies indicates the different sources of the laser-induced femtosecond photocurrents in a Pt/FeRh interface responsible for these signals.

To conclude, we reveal different spin-dependent sources of laser-induced THz emission in a Pt/FeRh bilayer. The most intense source originates from ultrafast laser-induced demagnetization of ferromagnetic FeRh and the inverse spin-Hall effect in the Pt-film. The second, weaker, source of THz emission is due to photo-induced magnetization tilting and inverse spin-orbit torque at the Pt/FeRh interface.

Support for this research was from the U. S. Department of Energy Grant No. DE-SC0018237, the Nederlandse Organisatie voor Wetenschappelijk Onderzoek (NWO), the European Union Horizon 2020 and innovation program under the FET-Open grand agreement no.713481 (SPICE), the European Research Council ERC grant agreement no.339813 (EXCHANGE), and grant agreement 852050 (MAGSHAKE) is greatly acknowledged. Authors want to thank Ray Descoteaux from CMRR, and Sergey Semin, Tonnie Toonen, and Chris Berkhout from Radboud University for all the technical support.

.

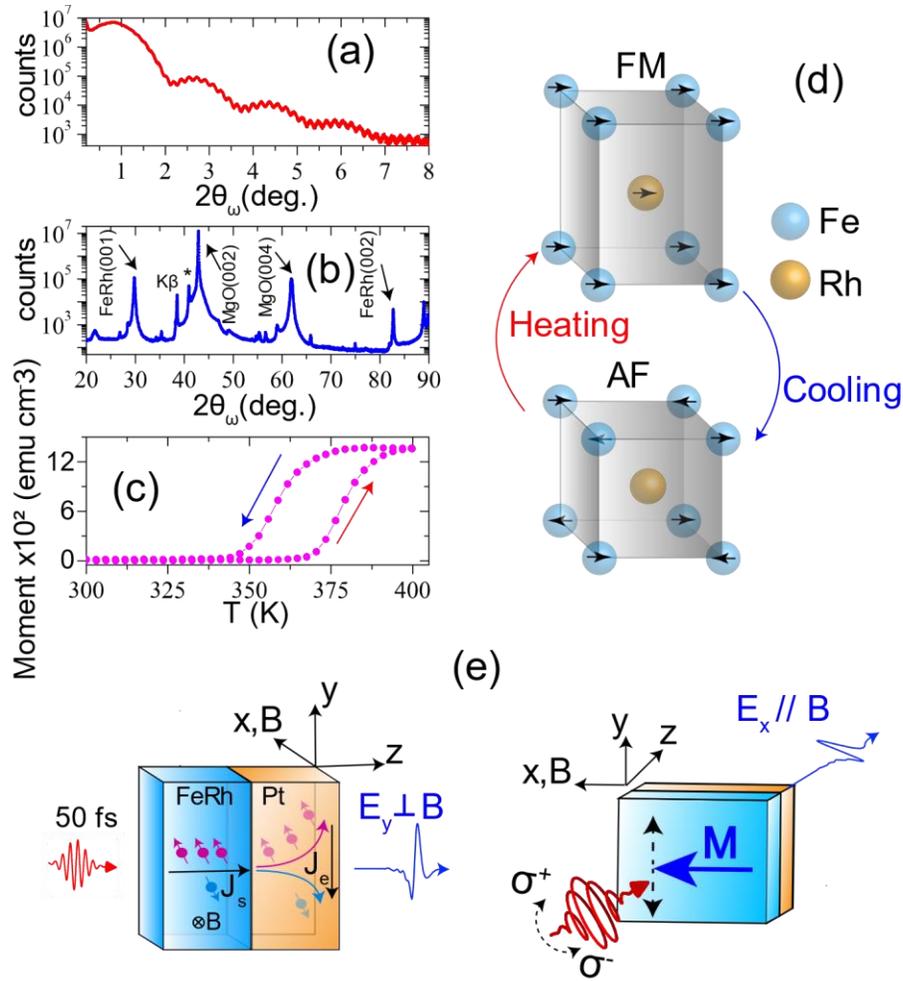

Figure 1 (a) X-ray reflectometry and (b) x-ray diffraction spectra measured in a 40 nm thick FeRh(001) sample deposited on a MgO(001) single crystal substrate and capped with a 5 nm thick Pt-layer. (c) The net magnetic moment of the 40 nm FeRh sample measured with a magnetic field strength of 150 mT applied in the plane of the sample. The AF-FM transition is hysteretic with temperature. (d) Illustration of magnetic and structural changes from AF to FM in the FCC-tetragonal FeRh unit cell. Upon heating the crystal lattice expands along the c-axis which results in an overall volume expansion by 1%. (e) Illustrations of the laser-induced femtosecond photocurrents generated by (left) the inverse spin Hall effect in Pt (vertical spin-pumping) and (right) the inverse of spin-orbit torque.

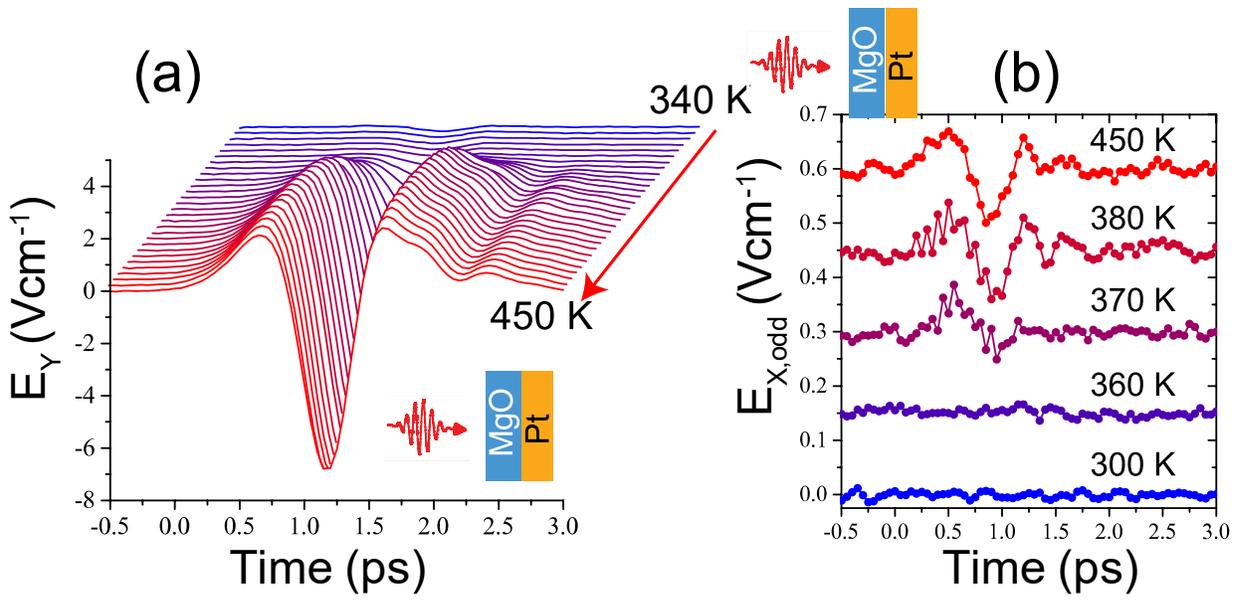

Figure 2(a) Time-traces of the helicity independent component, $E_Y$ of the laser induced electric field are shown for various initial temperature of the sample between 300-450 K. (b) same as (a) but for the helicity-dependent electric field component, $E_{X,odd}$.

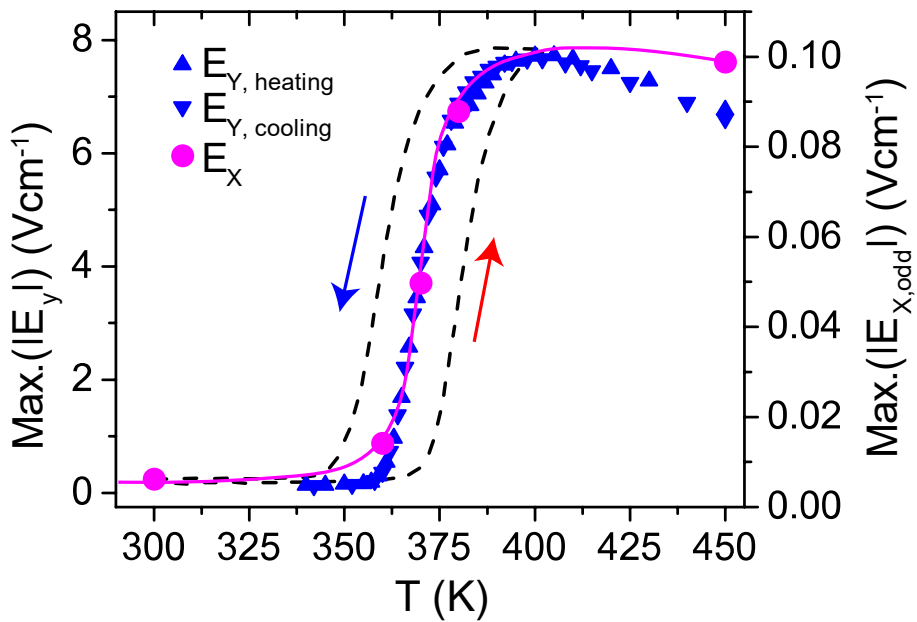

Figure 3 The peak amplitudes of the THz electric field for both $E_X$ and $E_Y$ are plotted as a function of temperature. Note that the Max.($|E_Y|$) are shown for both sample heating and cooling processes. The magenta solid curve is a guide to the eye for the Max.($|E_X|$) vs T data. For comparison the magnetic moments measured with VSM are shown as dashed black curves. The red and blue arrows correspond to the heating and cooling processes.

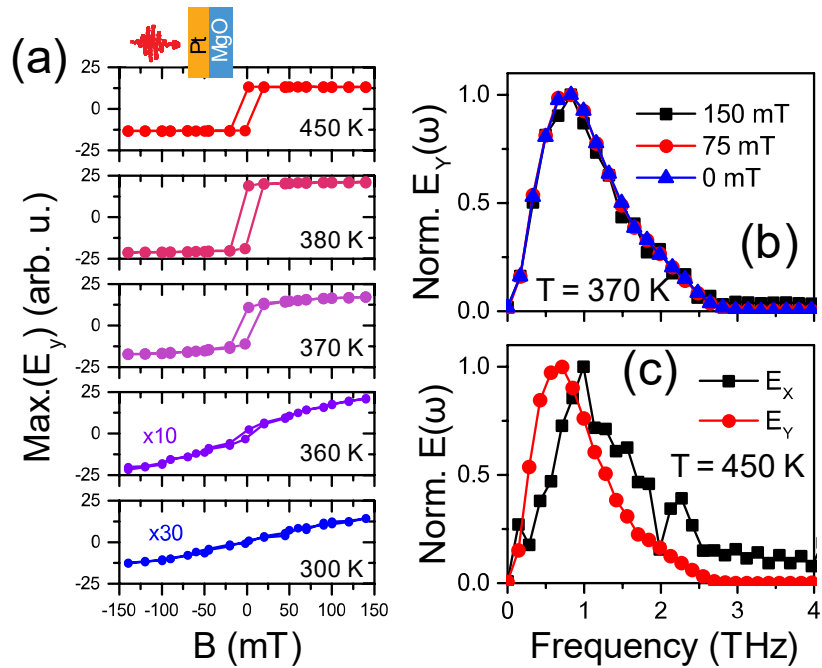

Figure 4 (a) The *Max.(|E_y|)* as a function of the magnetic field for various initial sample temperatures between 300 and 450 K. Note that the measurements were done along the cooling process (450 K-300 K). (b) Fourier spectra of the THz waveforms are shown for three different external magnetic field strengths detected at an initial sample temperature, T = 370 K. (c) The comparison of Fourier spectra for $E_X$ and $E_Y$ components of the THz electric field detected at T = 450 K for an applied magnetic field strength of 150 mT.

# Supplementary Materials for "Femtosecond Photocurrents at the Pt/FeRh Interface"

R. Medapalli[1,2], G. Li[3], Sheena K.K. Patel[1], R. V. Mikhaylovskiy[3,4], Th. Rasing[3], A. V. Kimel[3], and E. E. Fullerton[1]

[1]Center for Memory and Recording Research, University of California, San Diego, La Jolla, California 92093-0401, USA

[2]Department of Physics, School of Sciences, National Institute of Technology, Andhra Pradesh-534102, India

[3]Radboud University, Institute for Molecules and Materials, Heyendaalseweg 135, Nijmegen, The Netherlands

[4]Department of Physics, Lancaster University, Bailrigg, Lancaster LA1 4YW, UK

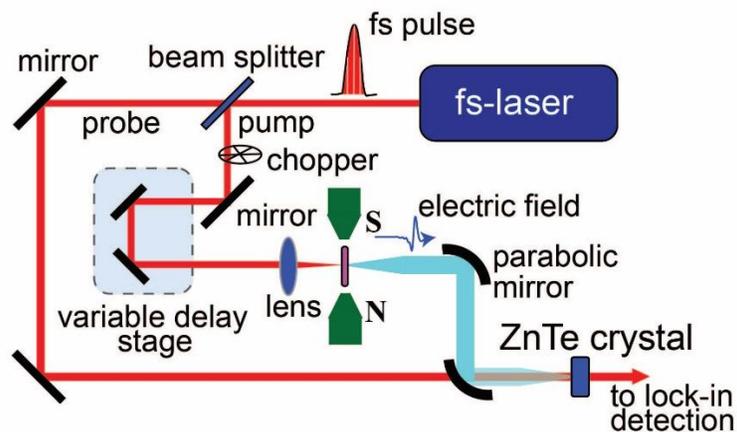

Supplementary Figure1. Schematic of the laser-induced THz detection technique is shown. Experimental scheme, consisting of two beams (λ = 800 nm), one for excitation of the sample and one for the sampling of the emitted THz electric field radiation. The sampling is based on the Pockels effect in a ZnTe crystal, in which the THz electric field pulse modulates the polarization state of the co-propagating sampling pulse. The change in the polarization state is detected by measuring the relative intensity of the two transverse sample electric field components. A combination of two wire-grid polarizers is used to detect, separately, the $E_X$ and $E_Y$ components of the laser-induced electric field. Our spectrometer is sensitive up to about 5 THz.

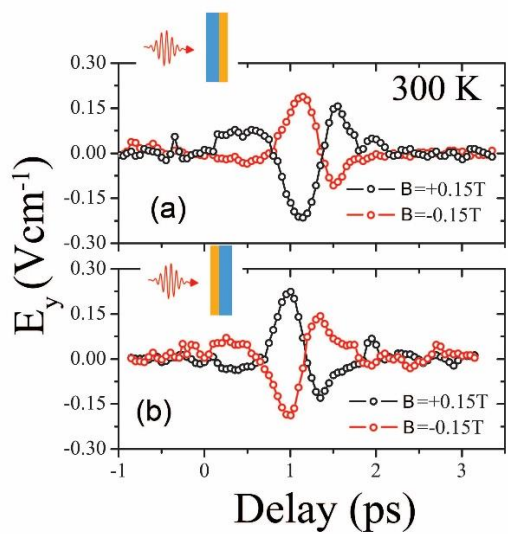

Supplementary Figure2 Time-traces of the THz signal measured at 300 K for two opposite orientations of the magnetization, $M↓$ (red) and $M↑$ (black) when the laser pumping was done from the MgO-substrate side (a) or from the Pt-cap side (b).

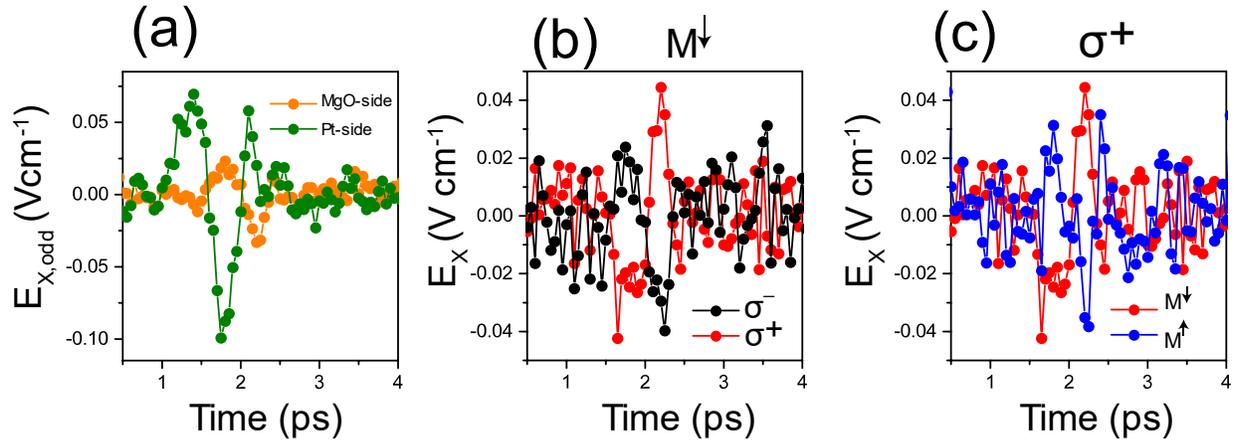

Supplementary Figure 3 (a) Time-traces of the helicity-odd THz electric field component $E_{X,odd}$ measured when the laser pumping was done from the MgO-substrate side (orange) and Pt-cap side (olive). (b) Time-traces of the $E_X$-component of the electric field for two different opposite pump helicities $\sigma^-$ (black) and $\sigma^+$ (red) for a given initial orientation of the magnetization, $M\downarrow$ (c) Time-traces of the $E_X$ component of the electric field for two opposite orientations of the magnetization, $M\downarrow$ (red) and $M\uparrow$ (blue) when pumped with right circularly polarized light ($\sigma^+$). The position of zero time is chosen arbitrarily but kept consistent between the measurements.